\RequirePackage{fix-cm}
\documentclass[smallextended]{svjour3}       % onecolumn (second format)
\smartqed  % flush right qed marks, e.g. at end of proof
\usepackage{graphicx}
\usepackage{amsmath}
\usepackage{amssymb}
\usepackage[subnum]{cases}
\usepackage{multirow,booktabs}
\usepackage[linesnumbered,ruled,vlined]{algorithm2e}
\usepackage{color}
\usepackage[font={small}]{caption}
\usepackage{subfig}
\usepackage{comment}
\usepackage[labelfont=bf]{caption}
\usepackage{url}
\begin{document}

%\title{WITS: An Internet of Things Aware, Personalized In-Home Monitoring Through the Web}

\title{Up in the Air: When Smart Homes Meet Internet of Things}


\author{Lina Yao, Quan Z. Sheng, Boualem Benatallah, Schahram Dustdar, Xianzhi Wang, Ali Shemshadi and Anne. H. H. Ngu\and
}


\institute{Lina Yao \at
              School of Computer Science and Engineering\\
              \email{lina.yao@unsw.edu.au}           %  \\
           \and
}


\maketitle

\begin{abstract}
Over the past few
years, activity recognition techniques have attracted unprecedented attentions. 
%
%Many applications reveal a special interest is in the pervasive e-Health domain, 
Along with the recent prevalence of pervasive e-Health in various applications such as smart homes, automatic activity recognition is being implemented increasingly for rehabilitation systems, chronic disease management, and monitoring the elderly for their 
%, as well as in 
personal well-being.
% applications. 
In this paper, we present WITS, an end-to-end web-based in-home monitoring system for convenient and efficient care delivery. The core components consist of a novel shared-structure dictionary learning approach combined with rule-based reasoning for continuous daily activity tracking and abnormal activities detection.
%\textcolor{blue}{:} 
WITS also exploits 
%we develop 
an Internet of Things (IoT) middleware 
for 
%providing a 
the scalable and seamless
% integrated framework for 
management and learning of the information 
%naturally 
produced by ambient sensors.
% with better interoperability and accessibility; 
%in addition, 
We further develop a user-friendly interface, which runs on both iOS and Andriod, as well as in Chrome, for the efficient customization of WITS monitoring services without programming efforts. 
%Herein we 
This paper presents the architectural design of WITS, the core algorithms, along with our solutions to the technical challenges in the system implementation. 
\end{abstract}

\vspace{-2mm}
\section{Introduction}
\label{Introduction}
The world population is aging rapidly due to the increasing life expectancy and declining birth rate.  The global share of elder people (aged 60 years or over) has increased from 9.2 percent in 1990 to 11.7 percent in 2013. It is expected that this trend will continue, reportedly reaching 21.1 per cent by 2050\footnote{\url{http://www.un.org/en/development/desa/population/publications/pdf/ageing/WorldPopulationAgeing2013.pdf}}. In the meantime, the human lifespan has increased as a result of increased health awareness and improved quality of food and medicine. Yet, the elderly are susceptible to various types of injuries and accidents and consequently require more medical care facilities. 
%And 
The increasing expenditures on health care for the elderly result in great financial impacts, urging healthcare providers to develop novel technologies with good usability and cost-effectiveness, such as smart homes, to address issues like monitoring the status of the elderly or chronically ill patients in their own homes.
Such techqnologies are essential to the elderly for them to live independently and safely \cite{mennicken2014today}\cite{dawadi2013automated}\cite{tung2013everyday}. 

A typical smart home is equipped with sensors, wherein cameras, IR sensors, ambient sound, heat, as well as contact sensors are mounted on furnitures and used in the home environment in a non-intrusive manner. The sensors continuously collect data about the location and activity of the subjects in the environment  without interfering with residents' daily activities. Such data would be a valuable asset to understand people's behaviors and their well-beings \cite{minor2015data}\cite{cook2015analyzing}\cite{bhattacharya2014using}. For example, in the following scenario, by monitoring the daily behavior routine of an old person, an assistant service can track how completely and consistently his daily routine is performed and on this basis to determine whether and when intervention or assistance is needed, e.g., whether the person is falling in bathroom, whether the person frequently forgets to take medicine or turn off the oven (which are early signs of dementia that harms his cognitive health), or whether the person has spent too much watching TV instead of doing exercises.

Over the years, significant research efforts have been contributed to developing smart home systems for monitoring the activities of daily living (ADLs). These efforts can be used to infer broader patterns such as common daily routines. However, several challenges remain concerning the interoperability among sensors and heterogeneous information, affordability, and accuracy. The rising of the Internet of Things (IoT) pushes the ambient intelligence forward by expanding the scale and scope of the healthcare domain significantly. IoT also bring new opportunities by facilitating the collection and reflection of diverse information in the physical world.

We propose a novel Web-based
%customized 
personal-wellness monitoring system to observe and quantify the residents' activities and abnormalities at homes. The proposed project is based upon our previous work on Internet of Things-aware smart homes \cite{yao2015web,yao2016context}, one of the stepping stones along a pathway of innovations that enable off-site specialists to observe and diagnose patients or residents' wellness reliably and accurately in real-time. It is especially important for special groups of people, such as old adults with physical and cognitive limitations or postpartum women, for whom it is often inconvenient or impossible to take outdoor activities.
%, and spend most of their time at home.

In this paper, we present WITS, \underline{W}eb-based \underline{I}nternet of \underline{T}hings \underline{S}mart home,
% application, 
as an end-to-end solution to facilitate the development of smart home applications. We propose to develop a hybrid recognition framework, which leverages multi-task learning, dictionary learning, and rule-based reasoning to observe and quantify the changes in the readings of sensors deployed in home environments, so as to continuously track residents' daily behaviors and detect any abnormal events for early and timely medical assistance. We develop a scalable IoT middleware that connects the physical world and the cyber world to overcome the {home interactivity} and interoperability issues. 

\begin{comment}
to enable always-on smart home monitoring. we design an unobtrusive, cost-effective radio frequency-based system wherein sensors are deployed in environments to infer people's movements or activities by analyzing received signal strength fluctuations with developing a set of data mining methods \cite{minor2015data, hong2013ambient, yaoicdm2015}. 
%(3) 
To support user-friendly programming, we develop a Web-based Trigger-Action programming model in a Drag-and-Drop style, which provides better usability and enables better practicality of allowing end users to customize home automation without programming efforts. 
%
\end{comment}

%\subsection{Challenges and Contributions}
WITS makes contributions by addressing three major challenges. 
%\textcolor{red}{new contributions for icdm, 
%1. activity-dictionary, and backward abnormal event detection 2. rule-based complex activity reasoning: The proposed method emphasizes the capability
%in dealing with the heterogeneity and uncertainty
%of context data. It also enables the system to learn by
%taking advantage of the recorded history context to
%derive underlying knowledge to supervise the service
%customization. 3. simply mention iot and tap programming}
%\begin{enumerate}

%\item 
%\vspace{2mm}
%\noindent{\em How to 
%%achieve 
%seamlessly integrate the heterogeneous contexts of a smart home with better interoperability and affordability?} 
%\begin{itemize}
%\item{\em 

\vspace{2mm}
\noindent{\bf How to seamlessly integrate the heterogeneous contexts of a smart home with better interoperability?}
There have been 
%a lot of 
many insights
% and directions 
proposed to boost smart home systems \cite{mennicken2014today}. However, most of current systems are still hampered by interoperability issues \cite{edwards2001home}. 
%Very few of them addressed {\em interoperability} well. 
We propose to address this challenge by developing an IoT middleware, which provides the necessary infrastructure to transparently and seamlessly glue heterogeneous resources and services together, by accessing sensors and actuators across different protocols, platforms, and locations over the Internet stack. 
%
%
% We develop an IoT middleware, wherein propose to leverage the power of Internet of Things,
%% (Section \ref{sec:imple}), 
%which relaxes users from the complicated process of sensor selection and enables them to better interpret the collected data - since possible use cases can be determined and programmed in apriori and to interface with many third-party applications, usually through some cloud storage services. 
Our IoT middleware 
%is  converging to 
makes smart home interactivity a reality, and enables smart homes to become more affordable and accessible with better interoperability between IoT-enabled objects, user devices, and cloud services by leveraging the Web architecture. Compared to existing IoT middlewares \cite{welbourne2009building}\cite{taylor2013farming}\cite{pintus2012paraimpu}, our work not only deals with the interoperability issue as many smart home systems but also offers an effective way to describe diverse objects for provisioning disparate information flow in smart home systems. 
%I also provides suitable integrated abstractions for heterogeneous events and information, which are converted to universal services, and can be easily invoked.
% by internal and external programs or service providers on the Web. 

%%\item
\vspace{2mm}
\noindent{\bf How to effectively recognize user activities and detect abnormalities by 
%effectively 
utilizing easily-accessible sensory data streams?}  
%
%Radio Frequency Identification Technology (RFID) is 
%%shaping up to be 
%an important building block for IoT \cite{welbourne2009building}
%%. We have seen 
%and more and more RFID 
%%tags and 
%devices are deployed in indoor settings. It is obviously a natural choice for us to make use of RFID signals in our design. 
%
%michael not sure the following make any sense.
%Smart environment algorithms need to recognize and track activities that people normally perform as part of their daily routines. To function independently at home, individuals need to be able to complete Activities of Daily Living such as sleeping, eating, grooming, cooking, drinking, and taking medicine etc.
%
The proposed system employs a series of core algorithms to recognize and track people's daily routines of activities and to further identify abnormalities. The objective is to discern activity patterns from the home sensory data by considering multidimensional contexts. 
However, 
%due to the following challenges, 
it has not been fully investigated how to interprete human behaviors from the massive, user-generated, heterogeneous, multi-modal, and context-relevant sensory data \cite{bulling2014tutorial}\cite{minor2015data}\cite{chen2012sensor}\cite{yao2016learning}, as a result of several challenging issues.
%
%\noindent {\em 1. 
The first issue is how to deal with intra-class activity variability. Since every people may have their  distinct behavior patterns, the same activity may be performed differently by different people. In many cases, the same person may perform the same activity differently, due to various factors such as the stress, fatigue, or emotion of the person, or even the surrounding environment.
The second issue is 
%\noindent {\em 2. How 
how to handle unseen samples effectively. It is usually hard to obtain a large set of activity behaviors owing to the difficulty of gathering sufficient training samples from multi-domain contexts (time-consuming and tedious annotating). 
The third challenge is that 
%\noindent {\em 3. How to reduce the need for labeled data. 
most approaches require intensive training and assume all of the training samples are annotated and available in advance. Yet such assumptions are often impractical for real-world applications, as in many cases, unlabeled new data keeps coming continuously.

To 
%attack 
tackle above challenges, we propose a novel method that leverages the joint advantages of multi-task learning \cite{evgeniou2007multi} and dictionary learning \cite{aharon2006img} for automatic activity recognition. Both of thea bove techniques have proven successful in a wide range of areas \cite{guha2012learning}\cite{yan2014multitask}\cite{wang2013multi}. Dictionary learning can generate a compact and discriminative manifesto of features, and multi-task learning can achieve good performance by learning multiple tasks that share commonalities simultaneously. Since such learning methods generally require more ground-truth data than we can obtain for complex activity inference, we introduce rule-based reasoning to help handle the unseen data. 
We specifically propose a joint framework to derive both activity-specific dictionaries (i.e., people-independent knowledge) and person-specific dictionaries, where intrinsic relationships
have invariant properties and are less sensitive and variant with different subjects.
Such properties can be used as signatures to profile activities and to support more accurate activity recognition 
%in recognition to tackle intra-class variability. We 
%further 
via rule-based reasoning based on multi-type contextual information (e.g., symbolic location and object use) . 

%
%We also design a bilinear SVM based localization algorithm, in which sensor signal and timestamps are formed as a two-dimensional matrix. The localization is developed to detect people presence to assist context-aware activity recognition, such as if a person is detected to fall, we can identify where he falls, e.g., in toilet or bathroom. 
%The proposed method shows more robust results compared with sliding window based algorithms

%\item 
\vspace{2mm}
\noindent{\bf How to  manage the personalized smart home in a more friendly way?} 
Another key challenge to smart homes is how to create customizable services in an effective and user-friendly manner. This involves designing an easy use and powerful interface for accessing, exchanging, and manipulating information from the smart homes, which is what most of the existing smart home systems lack, to pursue the user satisfactions.
%, where the residents interact with the environment and move through. 
%
% either lack of an interface for user-friendly interactions with smart homes
%, such as service customization or are difficult to use with limited functions. 
%or support simple rule specifications.

To address this challenge, we develop a user-friendly graphical interface using the Trigger-Action Programming (TAP) model to achieve value-added service customization. This model allows users to configure higher-level rules via a Web browser by integrating the inferred contexts, e.g., user locations, activities and objects in use, without programming effort. Various services and their semantic information are abstracted as universal services in WITS 
%in our IoT paradigm, 
and represented as icons on the graphical interface. Through this easy-to-use Web interface,
% (Section \ref{sec:imple}), 
users can specify and manage complex rules and build their personalized smart home applications in a 
%at 
Drag-and-Drop manner. 
%ease by freely combining different icons. 
Compared to the existing efforts that adopt TAP \cite{sohn2003icap,ur2014practical}, WITS can support 
%complex 
advanced rule customization with better scalability and flexibility.
%, details will be elaborated in Section \ref{sec:high}. 
%\end{enumerate}
%\end{itemize}

The remainder of this paper is organized as follows. We first review the related work. Next, we overview the WITS system, followed by
% in Section \ref{sec:overview}, followed by system design in Section \ref{sec:design} 
the core algorithms for activity recognition. We then describe our system implementation including software, hardware, and interface design, and 
% in Section \ref{sec:imple}. We report extensive experimental results.
% in Section \ref{sec:experiment} and 
finally, we conclude the paper by pointing out possible future directions.

\begin{table*}[!tb]
\begin{scriptsize}
\centering
\caption{IoT Middleware Comparison}
\label{tab:middleware}
\begin{tabular}{|p{2cm}|p{2.3cm}|p{0.3cm}|p{0.3cm}|p{0.4cm}|p{0.4cm}|p{0.5cm}|p{0.4cm}|p{0.5cm}|p{0.3cm}|p{0.3cm}|p{0.80cm}|p{1.0cm}|p{1.1cm}|}
\hline
\multirow{2}{*}{\textbf{Categories}} & \multirow{2}{*}{\textbf{Prototypes}} & \multicolumn{7}{c|}{\textbf{Functionality}} & \multicolumn{2}{c|}{\textbf{Interface}} & \multicolumn{3}{c|}{\textbf{Implementation}} \\ \cline{3-14}
& & \textbf{S} & \textbf{P} & \textbf{RM} & \textbf{RE} & \textbf{CAR} & \textbf{SN} & \textbf{MSS} & \textbf{M} & \textbf{W} & \textbf{RDL} & \textbf{API} & \textbf{PL} \\ \hline
\textbf{Smart Energy}& Webnergy & -- & -- & -- & -- & -- & -- & -- & $\bullet$ & $\bullet$ & JSON XML & REST & Java \\ \hline
\textbf{Tagging Objects} & EPCIS Web Adapter & -- & -- & -- & -- & -- & -- & -- & -- & $\bullet$ & JSON XML & REST & -- \\ \hline
\multirow{6}{*}{\textbf{IoT Middleware}} & Zetta & -- & -- & $\bullet$ & -- & -- & -- & $\bullet$ & -- & $\bullet$ & -- & REST & -- \\ \cline{2-14}
& Kaa IoT Platform & -- & -- & $\bullet$ & -- & -- & -- & $\bullet$ & -- & -- & -- & -- & -- \\ \cline{2-14}
& w3c/web of things & $\bullet$ & $\bullet$ & -- & $\bullet$ & -- & -- & -- & $\bullet$ & $\bullet$ & JSON TDL LD & REST & -- \\ \cline{2-14}
& ThngSpeak & -- & -- & $\bullet$ & -- & $\bullet$ & -- & -- & -- & $\bullet$ & -- & REST & -- \\ \cline{2-14}
& EVRYTHNG & $\bullet$ & $\bullet$ & $\bullet$ & -- & -- & $\bullet$ & $\bullet$ & -- & $\bullet$ & JSON & REST & -- \\ \cline{2-14}
& OpenIoT & $\bullet$ & $\bullet$ & $\bullet$ & -- & -- & -- & $\bullet$ & -- & $\bullet$ & SSN Specification & -- & -- \\ \hline
\multirow{14}{*}{\textbf{SmartHomes}} & openHAB & -- & -- & -- & $\bullet$ & -- & -- & $\bullet$ & -- & -- & -- & -- & -- \\ \cline{2-14}
& Eclipse Smart Home & -- & -- & $\bullet$ & -- & -- & -- & $\bullet$ & $\bullet$ & $\bullet$ & -- & -- & Java \\ \cline{2-14}
& OpenDomo & $\bullet$ & $\bullet$ & $\bullet$ & $\bullet$ & -- & -- & $\bullet$ & -- & -- & -- & -- & -- \\ \cline{2-14}
& FreeDomotic & $\bullet$ & $\bullet$ & $\bullet$ & $\bullet$ & -- & -- & -- & -- & -- & -- & -- & -- \\ \cline{2-14}
& Calaos & -- & -- & $\bullet$ & -- & -- & -- & -- & $\bullet$ & $\bullet$ & -- & -- & -- \\ \cline{2-14}
& MisterHouse & -- & -- & -- & $\bullet$ & -- & -- & -- & -- & $\bullet$ & -- & -- & Perl \\ \cline{2-14}
& Wosh & -- & $\bullet$ & $\bullet$ & -- & -- & -- & $\bullet$ & -- & $\bullet$ & XML & SOA & C++ \\ \cline{2-14}
& IoTivity & $\bullet$ & $\bullet$ & -- & -- & -- & -- & $\bullet$ & $\bullet$ & $\bullet$ & JSON & REST & -- \\ \cline{2-14}
& RaZberry & -- & -- & -- & -- & -- & -- & -- & $\bullet$ & $\bullet$ & JSON & -- & JavaScript \\ \cline{2-14}
& The Things System & $\bullet$ & $\bullet$ & $\bullet$ & $\bullet$ & -- & -- & $\bullet$ & -- & $\bullet$ & JSON & -- & node.js \\ \cline{2-14}
& PrivateEyePi & $\bullet$ & $\bullet$ & $\bullet$ & $\bullet$ & -- & -- & $\bullet$ & $\bullet$ & $\bullet$ & -- & -- & Python \\ \cline{2-14}
& IoTSyS & -- & -- & $\bullet$ & -- & -- & -- & $\bullet$ & -- & $\bullet$ & -- & -- & Java \\ \cline{2-14}
& House\_n & -- & -- & -- & $\bullet$ & $\bullet$ & -- & $\bullet$ & -- & -- & -- & -- & -- \\ \cline{2-14}
& CASAS & -- & -- & -- & $\bullet$ & $\bullet$ & -- & $\bullet$ & -- & $\bullet$ & -- & -- & -- \\ \hline
{\bf Hybrid project}& {\bf WITS}& $\bullet$ & $\bullet$& $\bullet$& $\bullet$ & $\bullet$ & $\bullet$& $\bullet$ & $\bullet$ & $\bullet$ &JSON & REST & C\#/Java \\ \hline
\multicolumn{14}{|p{16.5cm}|}{{\bf S:} Security (authentication, https etc) \quad {\bf P:} Privacy (access control etc.) \quad {\bf RM:} Real-time Monitoring \quad {\bf RE:} Rule Engine \quad {\bf CAR:} Complex Activity Reasoning \quad {\bf MSS:} Multi-source Support \quad {\bf SN:} Social Network (Twitter etc.) \quad {\bf M:} Mobile interface \quad {\bf W:} Web interface \quad {\bf RDL:} Resource Description Language \quad {\bf PL:} Programming Language \quad {\bf API:}application programming interface \quad $\bullet$: Function enabled \quad --: Not Available/Not Applicable} \\ \hline
\end{tabular}
\end{scriptsize}
\vspace{-2mm}
\end{table*}

\section{Related Work}
%Our proposed system provides 
%an end-to-end solution of Web-based in-home monitoring services. We cross-compare our system with current existing smart home systems and Internet of Things (IoT) middleware support. 

Over the years, significant research efforts have been contributed to smart home systems such as the Aware Home~\cite{kidd1999aware}, House\_n in MIT~\cite{intille2006using}, CASAS~\cite{rashidi2011activity}, and Adaptive House~\cite{mozer2004lessons}.
All these efforts focus on people's direct interactions with the technologies and help to infer broad patterns such as common daily routines such as facilitating the study of different smart home technologies in more depth and in contexts that closely resemble the real-world domestic spaces. However, several challenges remain: 1) heavily reliance on people's involvement such as the wearing of battery-powered sensors, which may not be practical in real-world situations; 2) lacking a synthetic way of deploying ubiquitous available sensors and taking advantage of disparate services offered by smart homes; and 3) lacking a user-friendly interface for end users to access and create personal rules for 
%enabling 
smart home 
%context-aware 
automation.  %Our work provides a complete treatment corresponding to these tackles. 

In addition, most existing IoT middlewares allow users to add sensors as they desire and offer users tools (simple App or Web browser) to view the sensor-collected raw data. However, Neither of them is able to support smart home applications very well because they usually provided limited functionalities in terms of helping user develop complex functionality, interacting with other applications, or interpreting the data into reusable high level services. Some systems even limit end-users on the types and the numbers of sensors that they can use.%, although they provide them
% enables users the chance to interpret the collected data. 
%since possible use cases can be determined and programmed in apriori, and to interact with many third party applications, usually through some cloud storage services with bottlenecks. 
In order to achieve scalability and usability, our proposed system not only monitors the data collected from sensor in real-time but also automatically converts the data into actionable information using intelligent event generator or according to users predefined conditions. This enables IoT applications to be developed based on high-level contexts that are independent of low-level physical properties of the sensors or devices. A thorough comparison of 22 existing IoT middleware systems and research prototypes including our proposed WITS is shown in Table~\ref{tab:middleware}. %We cross compared our proposed system with 
%a collection of work relate to IoT middleware and/or smart home systems these efforts over Functionality, Interface, and Implementation. 
bidding adieu to 

\section{WITS Overview}
\label{sec:overview}

\begin{figure*}[!tb]
%\vspace{-2mm}
\begin{center}
%\begin{minipage}{8cm}
%\includegraphics[width=\linewidth]{figure/Arch.png}
%\centering{(a)}
%\end{minipage}
%\hspace{1cm}
%\begin{minipage}{8cm}
\includegraphics[width=\textwidth]{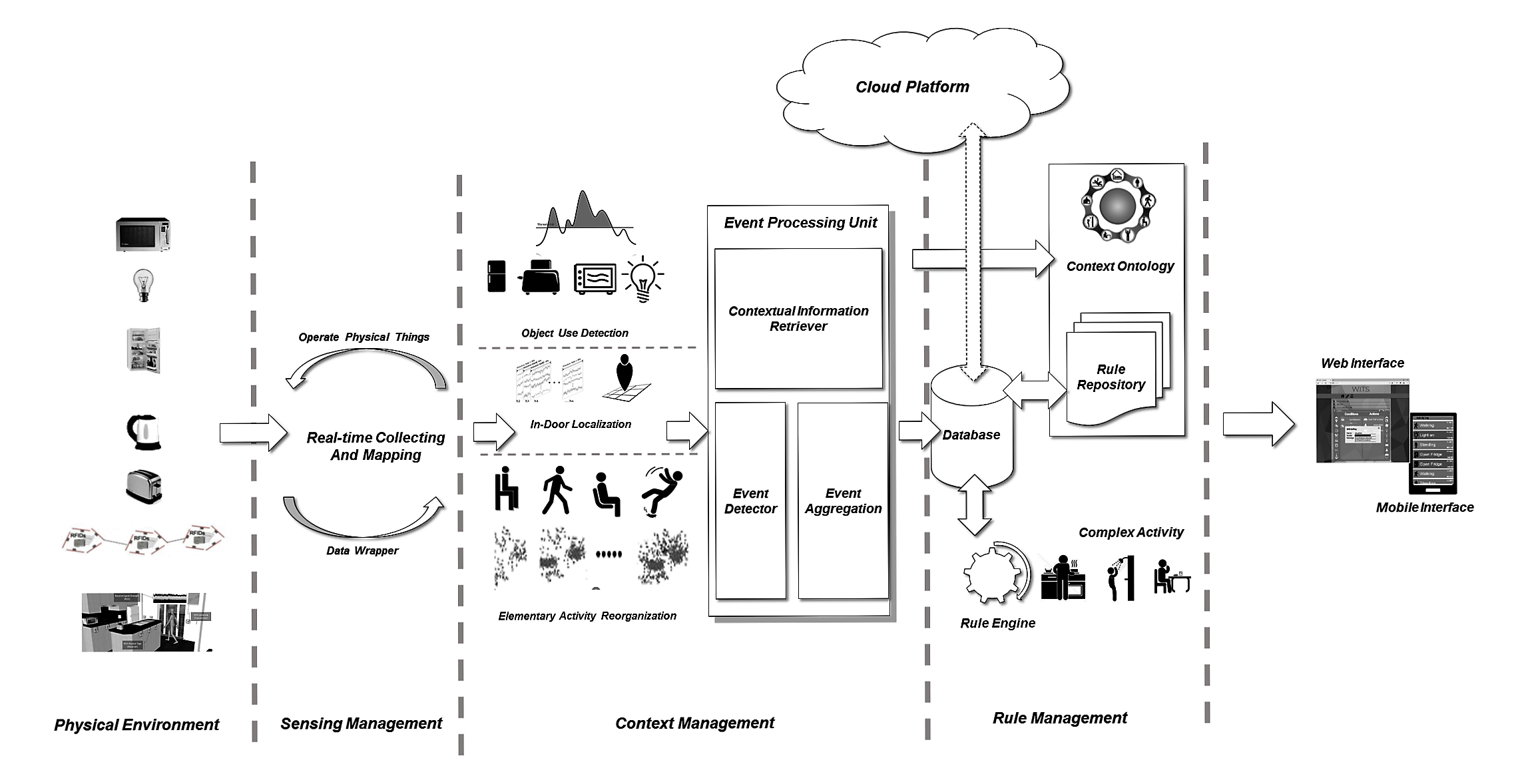}
%\centering{(b) }
%\end{minipage}
\caption{Overview of the WITS System Workflow}
%\label{fig:windowsize}
\label{fig:overview}
\end{center}
\vspace{-5mm}
\end{figure*}

%We first overview the WITS system architecture, followed by the description of the key components. 
%As depicted in , 
%workflow of 
WITS consists of three main 
%phases
components, namely {\em Sensing Management}, {\em Context Management}, and {\em Rule Management}, as shown in Figure~\ref{fig:overview}. 

The {\em Sensing Management} component manages all types of sensors (including RF tags and ambient sensors), 
collects and processes the raw data, and provides a universal API for higher-level programs to retrieve the status of physical entities. 
%In this phase, 
This component works in a scalable, plug-and-play fashion, where new sensors can be easily plugged-in and old sensors can be easily removed.
Several software components are 
%developed 
included in this part for filtering and cleaning raw sensor data, and adapting such data for high-level applications. Since some devices work with more than one sensor and the sensor readings may come asynchronously, a data access processor is used to allow the system to provide data synchronously, which lays the foundation of interoperability for smart home applications \cite{edwards2001home}.

WITS allows physical entities to be mapped to virtual resources to achieve a seamless integration of physical environment with organization business processes. % There are several languages such as Physical Markup Language\footnote{http://web.mit.edu/mecheng/pml}, Microformats\footnote{http://www.microformats.org}, and RDF (Resource Description Framework)\footnote{http://www.w3.org/RDF}, which can be used to add semantics to the descriptions of things.  
In WITS, we exploit the schema.org,
%\footnote{http://www.schema.org}, 
a recent initiative (launched in 2011 by Bing, Google, and Yahoo!) for different data semantics to be understood by both human users and machines. The universal RESTful API provided in our design enables higher level programs to retrieve the status of physical things with specified addresses, without needing to know where and how to find the physical sensors associated with the objects. 

This module is also in charge of data preprocessing. The physical activity and movement sensory data may contain noise, which deteriorates the clustering quality. To alleviate the disturbance from noise, we adopt the Hodrick-Prescott filter \cite{hodrick1997postwar}, a well-known trend analysis method in economics, to smooth the sensory data stream. This filter separates the time-series RSSI data into growth component $g_{t}$ and cyclical components $c_{t}$. The objective function is 
\begin{equation}
\sum_{t=1}^{T}c_{t}^{2}+ \lambda \sum_{t=2}^{T-1}((g_{t+1} - g_{t}) - (g_{t}- g_{t-1}))^{2}
\end{equation}
where $\lambda$ is the smoothing parameter. The programming problem is to minimize the objective over all $g_{t}|_1^{T}$.  
The incoming sensor readings are divided into data segments of length $\delta t$= 10 seconds. This time slice duration is long enough to be discriminative yet still short enough to provide accurate labeling results. The information of each segment is then transformed into 12 types
of statistical features such as Min, Max, Mean, Root Mean Square, Variance, Standard Deviation, Kurtosis, Skewness, Entropy, Median, Zero Crossing Rate, and Mean Cross Rate.

%
%The conceptual basis for this programming problem is that the first sum minimizes the difference between the data and its growth 
%component (which is the cyclical component) and the second sum minimizes the second-order difference of the growth component, which is analogous to minimization of the second derivative of the growth component.

%%% lina: remove for saving space %%%%
\begin{comment}
Figure~\ref{fig:hpfilter} shows that the cyclical component and the difference of a tag's signal variations when the resident is walking around. Bigger $\lambda$ ensure a smoothy difference but might trade off more information loss. We only use the growth component for further implementation. 
\begin{figure}[!tbh]
\vspace{-2mm}
\begin{center}
\includegraphics[width=6cm]{figure/hpfilter.eps}
\caption{}
\label{fig:hpfilter}
\end{center}
\vspace{-4mm}
\end{figure}
\end{comment}
%%%%%%%%%%%%%%%%%%%%%%%%%

%The {\em Event Management} 
The {\em Context Management} component aims at capturing contextual events. A typical smart home scenario involves three {\em atomic} contexts, namely basic human activities (e.g., getting up, walking, or lying on the bed), indoor localization (e.g., in bedroom or kitchen), and human-object interactions (e.g., using kettle or turning on light). Accordingly, a Event Detector employs a set of machine learning algorithms to analyze the signal fluctuations, in order to detect three contextual events,  
namely user activities,
%\ref{sec:dictionary}, 
symbolic location information (room-level position),
% \ref{sec:localization}, 
and object use events.
% \ref{sec:object}, 
%
%We develop several machine learning algorithms to interpret the signals, which will be elaborated in this paper later.
% following sections. 

%This phase 
WITS focuses on extracting and aggregating contextual events in a pipeline manner automatically. It extracts contextual information (e.g., temporal and spatial information), and then indexes and stores all the events with their related information in a database for data mining purposes. 
%A list of elements is constructed, storing the identifiers of objects, their types, and values, as well as the calculated contextual information. In this way, 
In WITS, users can focus on the functionalities of applications without worrying about issues such as connecting to the database, opening connections, querying with specified languages, or handling the results.
% (normally they are raw data and inconvenient to access). 
%
%One feature of WITS is that 
%it 
%can naturally harness the power of the Cloud platform. The events can be synchronized in a cloud platform for easy share.
% to support more user-centric rule customization. 
For example, in the context of aged care, doctors can access various personal data like pulse rate, blood pressure of a old person, whether the old person is taking medicine on time, or whether the person stays in a certain place for too long to make better decisions
%and 
%Such information will help the doctor . 
%It could be also subscribed and invoked from other public service providers, like weather forecast service or online traffic coordination system. 

The {\em Rule Management} component supports complex activity reasoning with the {\em Rule Engine} and high-level customizable composition to realize context-aware home automation. It derives underlying knowledge from the heterogeneous and uncertain historical context to supervise the service customization and
%, where  stores the generated rules. 
%The rules created in the rule generation can then be added to the rule library to supervise the context-aware automation service. 
enables the system to generate new context-aware automation rules (stored in the {\em Rule Repository}) and to adapt to dynamic context environments. In particular, an ontology-based knowledge base, {\em Context Ontology}, is used to represent the key contexts and their interrelations in smart home environments. It provides the formal semantics to describe context knowledge about objects, relationships and domain constraints, and is assisted with rule matching to support the sharing and integration of structured context information.

\subsection{Activity Recognition}
\begin{comment}
\subsubsection{Problem Description}
%In this section, we focus on introducing multi-task dictionary learning based activity recognition and anomaly detection. 
Let $\mathcal{S} \subset \mathbb{R}^{t}$ ($t$ is the number of ambient sensors deployed) be the domain of observable signal strength fluctuations of ambient sensors deployed in the house $\mathbf{s}$, and $\mathcal{L} \in \{1,...,K\} \subset \mathbb{R}$ be the domain of output activity label $l$ ($K$ is the number of activities) . Suppose we have $n$ signal and activity label pairs $\{(\mathbf{s}_{i},l_{i})| \mathbf{s}_{i}\in \mathcal{S}, l_{i}\in \mathcal{L}, i=1,...,n\}$. The training dataset
%would
can be expressed as:
\begin{equation}
\begin{split}
\mathbf{S}=[\mathbf{s}_1,...,\mathbf{s}_{n}]\in \mathbb{R}^{t\times n} \\
\mathbf{l}=[l_1,...,l_{n}]^{T}\in \mathbb{R}^{n}
\end{split}
\end{equation}
Our goal is to learn a predictor $\mathcal{F}: \mathcal{S} \rightarrow \mathcal{L}$ using the training dataset, to assign the most appropriate activity label for a given query sample.
\end{comment}
%
%\subsubsection{Model Formulation}
This section details the core techniques and algorithms for activity recognition in WITS. 

\subsubsection{Model Formulation}
Intuitively, it is reasonable to assume some
%underlying 
local commonalities under the intra-class variabilities
shared by all the activities. For example,
the same activities performed by different people should share some
commonality, e.g., walking forward and standing both shares
a torso perpendicular-like motion.
% or the same activities performed
%by the same person but with different time frames
%(e.g., in the morning or at night), while they hold differences
%inherited from different people. 
These intrinsic relationships
have invariant properties and are less sensitive to different people.
Thus, they can be used as a signature to
profile each activity style in activity recognition. Instead of uncovering the shared common space across activities using the original feature space or learning dictionaries from each activity separately, we aim at leveraging both the advantages of dictionary learning and multi-task learning by extracting the underlying $K$ style-specific dictionaries (person-independent), which are the more compact and discriminative representations of activities under a multi-task learning framework. We integrate them along with a collection of activity-specific dictionaries (person-dependent) in a joint framework

Let $\mathbf{X}_{k}\in \mathbb{R}^{n_{k}\times m} = [\mathbf{x}_{k}^{1}, \mathbf{x}_{k}^{2}, ... , \mathbf{x}_{k}^{n_{k}}]$ be the sensor samples, where $\mathbf{x}_{k}^{i} \in \mathbb{R}^{m}$ is a $m-$dimensional feature vector of sensory data, $n_{k}$ is the number of samples in the $k-$th task, and $k \in [1, K]$ tasks corresponding to $K$ activity styles. We assume a shared structure between different activities $\mathbf{Q} \in \mathcal{R}^{m \times sd}$, where $sd < m$ is the dimensionality of the subspace to project the original feature space into a low dimensional subspace. Furthermore, we consider the relationships between these weight vectors imposed by the neighbourhood structure of sensory data points that characterize each type of activity and define the following objective function:
%\begin{equation}
%\begin{split}
%\mathcal{J} = &\min_{\mathbf{D}_{k}, \mathbf{C}_{k}, \mathbf{P}, \mathbf{D}}\sum_{k=1}^{K}||\mathbf{X}_{k} - \mathbf{C}_{k}\mathbf{D}_{k}||_{F}^{2} + \lambda_{1}\sum_{k=1}^{K}||\mathbf{C}_{k}||_{1} \\&+ \lambda_{2}\sum_{k=1}^{K}\sum_{a,b}\mathbf{W}_{a,b}||\mathbf{C}_{k}(a,:) - \mathbf{C}_{k}(b,:)||_{F}^{2} \\
%&+ \lambda_{3}\sum_{k=1}^{K}||\mathbf{X}_{k}\mathbf{P} - \mathbf{C}_{k}\mathbf{D}||_{F}^{2} \\
%&\text{s.t.} \mathbf{P}^{\top}\mathbf{P} = \mathbf{I} \\
%& (\mathbf{D}_{k})_{j.}(\mathbf{D}_{k})^{\top}_{j.} \neq 1, \forall j = 1,...,l \\
%& \mathbf{D}_{j.}\mathbf{D}_{j.}^{\top} \neq 1, \forall j = 1,...,l \\
%\end{split}
%\end{equation}
%
\begin{equation}
\begin{split}
\mathcal{J}(\mathbf{X}_{k}, \mathbf{C}_{k}, \mathbf{D}_{k}, \mathbf{D}, \mathbf{Q}) =
&\underbrace{\sum_{k=1}^{K}||\mathbf{X}_{k} - \mathbf{C}_{k}\mathbf{D}_{k}||_{F}^{2}}_{1} + \underbrace{\lambda_{1}\sum_{k=1}^{K}||\mathbf{C}_{k}||_{1}}_{2} \\&+ \underbrace{\lambda_{2}\sum_{k=1}^{K}\sum_{a,b}\mathbf{W}_{a,b}||(\mathbf{C}_{k})_{a.} - (\mathbf{C}_{k})_{b.}||_{F}^{2}}_{3} \\
&+ \underbrace{\lambda_{3}\sum_{k=1}^{K}||\mathbf{X}_{k}\mathbf{Q} - \mathbf{C}_{k}\mathbf{D}||_{F}^{2}}_{4}\\
%&\text{s.t.} \mathbf{P}^{\top}\mathbf{P} = \mathbf{I} \\
%& (\mathbf{D}_{k})_{j.}(\mathbf{D}_{k})^{\top}_{j.} \neq 1, \forall j = 1,...,l \\
%& \mathbf{D}_{j.}\mathbf{D}_{j.}^{\top} \neq 1, \forall j = 1,...,l \\
\end{split}
\label{equ:overall}
\end{equation}
where $\mathbf{D}_{k} \in \mathbb{R}^{d \times m}$ is the activity dictionary ($d \le m$), 
%michael: seems not appear in the forumla.
%$(\mathbf{D}_{k})_{j.}$ in the constraints denotes the $j$-th row of $\mathbf{D}_{k}$, 
and $\mathbf{C}_{k}\in \mathbb{R}^{n_{k} \times l}$ corresponds to the sparse representation coefficients of $\mathbf{X}_{k}$. $\mathbf{D} \in \mathbb{R}^{d \times sd}$ is the activity style dictionary learned in the shared subspace and $\mathbf{D}_{j.}$ in the constraints denotes the $j$-th row of $\mathbf{D}$. There are four terms in the right side of the equation:

\vspace{2mm}
\noindent {\bf 1. Reconstruction Error on Individual Dictionary.}
The reconstruction error (the first term) for a usual/normal activity should be a small value because the learned dictionary represents knowledge in the previously seen sensory data. A small reconstruction error indicates the information within the newly observed data segment has appeared in early available samples. 

\vspace{2mm}
\noindent {\bf 2. Sparsity Regularization.} Since the dictionary is learned to maximize the sparsity of reconstruction vectors for normal activities (along with a fairly small reconstruction error), it is necessary to impose sparsity for reconstructing normal events. The sparsity regularization (the second term) ensures that reconstruction vectors are sparse for normal activities while dense for unusual/abnormal activities.

\vspace{2mm}
\noindent {\bf 3. Smoothness Regularization.}
The smoothness regularization (the third term) is calculated based on the fact that neighboring sensory variations are more likely to be involved in a same categorical behavior, where $\mathbf{W} \in \mathbb{R}^{n_{k} \times n_{k}}$ denotes the adjacency matrix of $[\mathbf{x}_{k}^{1},...,\newline \mathbf{x}_{k}^{n_{k}}]$.
We use the cosine similarity to measure the affinity. 

\begin{equation}
\mathbf{W}_{a,b} = \dfrac{\mathbf{C}_{k}(a,:)\cdot \mathbf{C}_{k}(b,:)}{\sqrt{\sum_{i}\mathbf{C}_{k}^{2}(a,i)}\sqrt{\sum_{i}\mathbf{C}_{k}^{2}(b,i)}}
\end{equation} 

\vspace{2mm}
\noindent {\bf 4. Reconstruction Error on Shared Dictionary.}  The reconstruction error of the shared dictionary (the fourth term) manifests the underlying commonalities between the people who perform the same activities with multi-task learning. The term is used to resist the noisy data and reduce the misclassification caused by inter-person variability. It should be small for normal activities, while big for abnormal activities. 

\subsubsection{Optimization}
The objective function (Equation~\ref{equ:overall}) quantifies the normality of activity in every data segment $\mathbf{X}_{k}$ with any reconstruction weight vector $\mathbf{C}_{k}$, any people-specific dictionary $\mathbf{D}_{k}$, and activity-specific dictionary $\mathbf{D}$. The lower $J$ is, the more likely $\mathbf{X}_{i}$ is generated by normal behavior. We obtain the optimal weight vectors $\mathbf{C}_{k}^{*}$ and dictionaries ($\mathbf{D}_{k}^{*}$, $\mathbf{D}^{*}$, $\mathbf{Q}^{*}$) by solving the following optimization problem:
\begin{equation}
\label{equ:opt1}
\begin{split}
(\mathbf{C}_{k}^{*}, \mathbf{D}^{*}, \mathbf{D}_{k}^{*}, \mathbf{Q}) = \arg \min_{\mathbf{D}_{k}, \mathbf{C}_{k}, \mathbf{Q}, \mathbf{D}} \sum_{k=1}^{K}J(\mathbf{X}_{k}, \mathbf{C}_{k}, \mathbf{D}, \mathbf{D}_{k}, \mathbf{Q})\\
\text{s.t.} \mathbf{Q}^{\top}\mathbf{Q} = \mathbf{I} \\
 (\mathbf{D}_{k})_{j.}(\mathbf{D}_{k})^{\top}_{j.} \neq 1, \forall j = 1,...,l \\
\mathbf{D}_{j.}\mathbf{D}_{j.}^{\top} \neq 1, \forall j = 1,...,l \\
\end{split}
\end{equation}
Since there exists $\ell_1$ minimization, we use alternate optimization algorithm to find a local optimum. In particular, we alternatively minimize one variable while fixing the other three parameters.
For example, with a learned dictionary $\mathbf{D}^{*}$, given a newly observed activity $\mathbf{X}'$, the algorithm learns the optimal reconstruction weight $\mathbf{C}'_{k}$, 
$\mathbf{Q}'$, and $\mathbf{D}'$. Consequently, $J(\mathbf{X}', \mathbf{C}'_{k}, \mathbf{Q}', \mathbf{D}^{*})$ measures the normality 
of $\mathbf{X}'$. $\mathbf{X}'$ is detected as unusual if its corresponding $J(\mathbf{X}', \mathbf{C}'_{k}, \mathbf{Q}', \mathbf{D}*)$ is larger than a certain threshold. 

\vspace{2mm}
\noindent{\bf Learning $\mathbf{D}$ with $\mathbf{D}_{k}$, $\mathbf{C}_{k}$ and $\mathbf{Qs}$ fixed.} With the dictionary $\mathbf{D}$ fixed, the optimization problem turns into:
\begin{equation}
\begin{split}
\min_{\mathbf{D}}\sum_{k-1}^{K}||\mathbf{X}_{k}\mathbf{Q} - \mathbf{C}\mathbf{D}||
%\text{s.t.}\hspace{4mm} \mathbf{D}_{j.}\mathbf{D}^{\top}_{j.} \neq 1, \forall j = 1,...,l
\text{s.t.} \hspace{3mm} \mathbf{D}_{j.}\mathbf{D}_{j.}^{\top} \leq 1, \forall j = 1,...,l \\
\end{split}
\label{equ:d}
\end{equation}
The constraint in Equation~\ref{equ:d} is to prevent the terms in $\mathbf{D}$ from being arbitrarily large, which would lead to 
%arbitrarily 
small values of coefficients. The above optimization problem is a least square problem with quadratic constraints, which can be solved using Lagrange dual. We adopt the efficient Algorithm 2 in \cite{mairal2009online}. 

\vspace{2mm}
\noindent{\bf Leaning $\mathbf{D}_{k}$ with $\mathbf{D}$, $\mathbf{C}_{k}$ and $\mathbf{Q}$ fixed.} To compute $\mathbf{D}_{k}$, the optimization problem is as follows.
\begin{equation}
\begin{split}
\min_{\mathbf{D}_{k}} ||\mathbf{X}_{k} - \mathbf{C}_{k}\mathbf{D}_{k}||_{F}^{2} 
\text{s.t.}(\mathbf{D}_{k})_{j.}(\mathbf{D}^{\top}_{k})_{j.} \leq 1, \forall j = 1,...,l
\end{split}
\end{equation}
This problem can also be solved in closed-form using the same algorithm \cite{mairal2009online}. 

\vspace{2mm}
\noindent{\bf Learning $\mathbf{Q}$ with $\mathbf{D}_{k}$, $\mathbf{C}_{k}$ and $\mathbf{D}$ fixed.} The optimization problem is 
as follows:
\begin{equation}
\begin{split}
\min_{\mathbf{P}} \sum_{k=1}^{K}||\mathbf{X}_{k}\mathbf{Q} - \mathbf{C}_{k}\mathbf{D}||_{F}^{2} 
\text{s.t.}\hspace{4mm} \mathbf{Q}^{\top}\mathbf{Q} = \mathbf{I}
\label{equ:p}
\end{split}
\end{equation}
We 
%can 
derive $\mathbf{D} = (\mathbf{C}^{\top}\mathbf{C})^{-1}\mathbf{C}^{\top}\mathbf{X}\mathbf{Q}$ according to the Lagrange dual and replace $\mathbf{D}$ in above Equation~\ref{equ:p}, which can be rewritten as:
\begin{equation}
\min_{\mathbf{P}} \sum_{k=1}^{K}tr(\mathbf{Q}^{\top}\mathbf{X}^{\top}(\mathbf{I} - \mathbf{C}(\mathbf{C}^{\top}\mathbf{C})^{-1}\mathbf{C}^{\top})\mathbf{X}_{k}\mathbf{Q})
\end{equation}
The optimal $\mathbf{P}$ is composed of eigenvectors of the matrix $\mathbf{X}^{\top}(\mathbf{I} - \mathbf{C}(\mathbf{C}^{\top}\mathbf{C})^{-1}\mathbf{C}^{\top})\mathbf{X}$ corresponding to the $s$ smallest eigenvectors. 

\vspace{2mm}
\noindent{\bf Learning $\mathbf{C}_{k}$ with $\mathbf{D}_{k}$, $\mathbf{Q}$ and $\mathbf{D}$ fixed.} Equation~\ref{equ:overall} is equivalent to the following optimization problem:
\begin{equation}
\begin{split}
&\min_{\mathbf{C}_{k}}\sum_{k=1}^{K}||\mathbf{X}_{k} - \mathbf{C}_{k}\mathbf{D}_{k}||_{F}^{2} + \lambda_{1}\sum_{k=1}^{K}||\mathbf{C}_{k}||_{1} \\&+ \lambda_{2}\sum_{k=1}^{K}\sum_{a,b}\mathbf{W}_{a,b}||(\mathbf{C}_{k})_{a.} - (\mathbf{C}_{k})_{b.}||_{F}^{2} \\
&+ \lambda_{3}\sum_{k=1}^{K}||\mathbf{X}_{k}\mathbf{Q} - \mathbf{C}_{k}\mathbf{D}||_{F}^{2} \\
%&\text{s.t.} \mathbf{P}^{\top}\mathbf{P} = \mathbf{I} \\
%& (\mathbf{D}_{k})_{j.}(\mathbf{D}_{k})^{\top}_{j.} \leq 1, \forall j = 1,...,l \\
%& \mathbf{D}_{j.}\mathbf{D}_{j.}^{\top} \leq 1, \forall j = 1,...,l \\
\end{split}
\end{equation}
As the above equation contains $\ell_1$ minimization, it is a non-convex problem. We adopt the feature-sign search algorithm \cite{lee2006efficient}, which considers the non-zero elements of coefficients to convert this objective function into a standard, unconstrained quadratic optimization problem, to solve the $\ell_1$ regularization problem. After obtaining the optimal dictionaries $\mathbf{D}^{*}$ and $\mathbf{D}_{k}^{*}$, given a test sample $\mathbf{X'}$, the activity label of $\mathbf{X}'$ can be obtained by computing its sparse coefficient $\mathbf{\hat{C}}_{k}$ and the minimal reconstruction error. 

In addition, the proposed method can detect unusual activities 
%which defined as unusual/rare activities in this work 
(e.g., falls) after learning the optimal behavior-specific dictionary and person-specific dictionary. Assume that we have instances of all the normal activities, then we can detect a test instance whose patterns deviate from the learned model as an anomaly as long as a activity label is assigned to the test instance.
For example, given a newly observed sensor data segment $\mathbf{X}'$ and current dictionaries $\mathbf{D}*$ and $\mathbf{D}_{k}^{*}$ and computed sparse coefficients $\mathbf{C}'_{k}$, a test instance is detected as unusual activity segment if the following criterion is satisfied:
$J(\mathbf{X}', \mathbf{C}', \mathbf{Q}', \mathbf{D}^{*}) > \epsilon$, where $\epsilon$ is a threshold defined by users.

\subsection{Coupling with Rule-based Reasoning}
\label{sec:high}
%As stated in the Section~\ref{sec:introduction}, 
Compared to basic activities 
%in Section~\ref{sec:activity} 
like walking, sitting and lying, which purely rely on signal fluctuations and 
%did not 
do not consider any contextual information (e.g., location information or object use), we combine such contextual information to derive a series of complex activities.
A complex activity is defined as a set of activities, multiple objects, and symbolic locations involved in a complex activity.
For example, cooking is a complex activity, including a series of actions like walking, 
%waving 
moving arms, standing etc, and a set of objects such as chop-board, knives, and oven.
% via 
Since cooking usually involves a large number of different objects which are shared across multiple activities, object usage information and symbolic information (e.g., kitchen or living room) can help discriminate between activities such as making a toast or making a coffee. Such distinctions can be important for applications 
%domains 
such as health monitoring or memory aids.
%
%
%As another example, 
%The Sleeping domain sleeping involves a series of rolling or trivial movements, with specific symbolic location information. 
%We can easily different lying 

Since it is often difficult, even impossible in some cases, to obtain sufficient training data with involving such a broad range of contexts,
%Moreover, obtaining such data, especially with sufficiently annotations or the ground truth, is tedious, time-consuming, error-prone, and may even be impossible in some cases (e.g., older people with dementia or collecting data from doing dangerous activities). 
we further develop a rule-based approach that leverages heterogeneous information such as object uses and location information (e.g., room presents) to recognize complex activities.
The rule-based method can also help handle the unexpected cases, such as the unexpected changes in a context or an unseen activity that arrives after training the model, and improve the performance in the activity recognition. We propose to couple shared structure learning with a rule-based method to provide a definitive solution to activity recognition. It has the following advantages: 1) capturing the explicit patterns without training efforts; 2) improving the overall performance by correcting the learning mistakes and handling unseen cases; 3) covering cases that a pure learning method cannot handle, e.g., inferring complex activities based on actions and resident's locations.

To 
%build a 
improve the usability of the system, we use the Trigger-Action Programming (TAP) programming paradigm for users to create rules.
TAP is a simple programming model applied in smart home applications, in which the user associates a trigger with an action, such that the action can be automatically executed when the trigger event occurs. The most popular TAP interface is an online service called if-this-then-that (IFTTT). IFTTT allows users to create programs that can automatically perform actions like sending alerts or changing settings of a smart home when certain events or conditions occur.

The {\em Visual Rule Programming Interface} component of WITS not only enables end-users to create rules for smart home automation but also offers a rule-based reasoning for high-level service (or resource) composition under Trigger-Action Programming paradigm through a Web-based rule editing interface. 
%Trigger-action programming (TAP) 
in our system, a rule consists of two parts: a {\em trigger} 
%{\em Events} 
and an {\em action}. 
A trigger is a composition of a set of boolean expressions. An action is a set of settings of entities. As shown in the sub figure of Figure~\ref{fig:interface}, \textit{Toilet.Occupied=true $\wedge$ Duration>= 30mins}, an action, \textit{SendAlert}, will be triggered to send an alert to the corresponding agent 
(e.g., a caregiver).
Compared to previous work that applies TAP in smart home applications, which only support a limited number of logical operators, our system has three advantages: i) it supports richer logical operators 
%such as multiple 
to help end-users creat more complex rules \cite{sohn2003icap},
ii) it supports a broader range of triggers such as human activities, changes of object status, and human mobility \cite{ur2014practical},
%3) thanks to IoT support, our system also can be naturally integrated into a Cloud platform for more advanced functionalities. 
and iii) it tolerates ambiguous status and events and allows flexible conjunctions of events and actions. All captured events can be treated as triggers or actions, leading to flexible rule customization.

\begin{table}[tb]
\begin{center}
\begin{scriptsize}
\caption{Illustrative Examples of Defined Rules}
\label{tab:activityrule}
\begin{tabular}{|p{0.85cm}|p{4cm}|p{2.5cm}|}
\hline
{\bf Index} & \textbf{Trigger} & \textbf{Actions} \\ \hline \hline
Rule 1 & \textit{Kitchen.Presence == True $\wedge$ Cooktop.ON == True $\wedge$ Choptable.Use == True} & Making Sandwich is detected \\ \hline
Rule 2 & \textit{Sleeping == True} & Turn off lights \\ \hline
Rule 3 & \textit{Toilet.Occupied == True $\wedge$ Duration $\geq$ 30mins} & Sending an alarm to caregiver \\ \hline
Rule 4 & \textit{Falling == True} & Sending an alarm to caregiver \\ \hline
Rule 5 & \textit{Porch.Presence == True $\wedge$ Time is {[}8:00pm 8:00am{]}} & Turn on front door light \\ \hline
Rule 6 & \textit{Couch.Occupied == True $\wedge$ TV.ON == True} & Watching TV \\ \hline
\end{tabular}
\end{scriptsize}
\vspace{-4mm}
\end{center}
\end{table}

\section{Implementation and Evaluation}
\label{sec:imple}
%This section presents both the hardware composition and software realization of WITS.
\subsection{Experimental Setup}
\begin{figure*}[!tbh]
%\vspace{-2mm}
\begin{center}
\includegraphics[width=\linewidth]{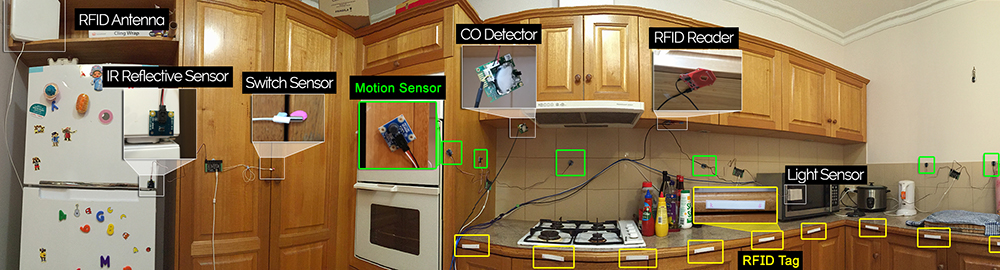}
\caption{Deployment of sensing units inside the kitchen area of the house}
%\label{fig:windowsize}
\label{fig:settingup}
\end{center}
\vspace{-5mm}
\end{figure*}

\noindent{\bf Hardware.} 
Since it is time-consuming to deploy a large number of tags and also not practical to tag
some objects because of their material (e.g., metal) or usage (e.g., objects used in a microwave oven), we propose to use only
some key objects for a specific set of activities by augmenting
the object usage with a complementary sensing technique
(i.e., accelerometers). Locations of tags include doors, cupboards,
refrigerator, dishwasher, etc (see Figure~\ref{fig:settingup}). 
The detailed configurations of the system like tag density, the shape of the sensor arrangement, sampling rate, have been thoroughly studied in our previous work \cite{yao2016learning,yao2015freedom}. 
%Due to space constraints and to be simplicity, we focus on the evaluation of the whole system in this paper. 

%%% old content  %%%
\begin{comment}
We use one Alien 9900+ RFID reader, four circular antenna (an antenna for each room) and multiple Squig inlay passive RFID tags in our experiment. The tags are placed along the walls, where each grid is roughly $0.8m \times 0.8m$.
% (Section~\ref{sec:tagdensity}). 
The antennas are arranged between $\approx 1.3m \sim 1.6m$ height with angle $\approx 70^\circ$. We attach the tags along the walls of each room, including 12 tags in the bedroom, 5 tags in the toilet, 10 tags in the bathroom, and 16 tags in the kitchen (Figure~\ref{fig:settingup}). The reader is connected to a backend server via an Ethernet cable and continuously collects sensor data. We also deploy a set of sensors (e.g., motion sensor, pressure sensor, light sensor).
% with objects using Arduino Yun and Phidget. 
\end{comment}

\vspace{2mm}
\noindent{\bf Data Processing.} The data collection and processing 
%and UI modules 
of WITS are fully implemented using the Microsoft .NET framework, and data are analyzed using MATLAB 2014a. The data collection module is developed with deploying based on the developer SDK and open APIs provided by Alien\footnote{http://www.alientechnology.com/}, Arduino\footnote{https://www.arduino.cc/}, Rasberry Pi\footnote{https://www.raspberrypi.org/} and Phidgets\footnote{http://www.phidgets.com/}. 

\begin{figure}[!tb]
%\vspace{-2mm}
\begin{center}
\includegraphics[width=0.8\linewidth]{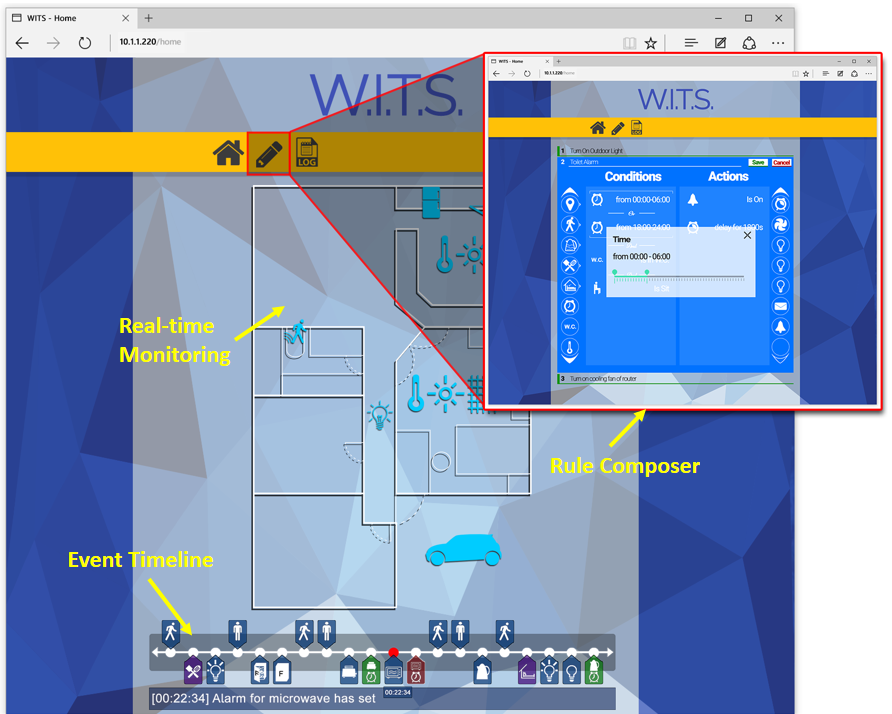}
\caption{WITS system: {\em Event Timeline} shows what is going on in the house, e.g., which room is the person in, what activity is performing and which object is used. The sub figure shows the {\em Rule Composer} interface.
% Rule composer interface: 
For example, for editing a rule like ``send an alarm when a person stays in toilet over 30 minutes''. A user only needs to drag the person, toilet and clock icons to the Trigger subpanel, the alarm icon to the Action subpanel and then performs some simple adjustments (e.g., adjust the clock slider to set the time period).}
%\label{fig:windowsize}
\label{fig:interface}
\end{center}
\vspace{-5mm}
\end{figure}

\vspace{2mm}
\noindent{\bf Graphical UI Modules.} WITS provides a holistic interface for the real-time monitoring, which is a scene-based graphical interface as illustrated in Figure \ref{fig:interface}. It works like an extended Harry Potter's Maurander's Map, wherein users can vividly observe what is happening. We also develop a graphical interface for rule customization. The detected events are associated with icons. Therefore, end-users only need to drag and drop the icons to set up their own personal rules (Figure \ref{fig:interface}). This Web interface represents a new direction in integrating information from both the physical and virtual worlds, which brings things, locations and activities together over a flexible IoT middleware, helping people be aware of their surrounding environments and thereby making better decisions. The practical experiences gained from this IoT framework provide insights into how IoT can be applied to support critical real-world applications such as assisted living of the elderly. 

\vspace{-2mm}
\subsection{Evaluation}
%In this section, we first report the evaluation results by comparing
%with the state-of-the-art activity recognition approaches using
%four publicly available real-world datasets.
%
%
We evaluate our approach by examining the accuracy of recognizing both low-level activities (inferred from signal fluctuations) and high-level activities (inferred from low-level activities, along with object usage and location information (Section~\ref{sec:high})). The detailed activities evaluated are summarized in Table~\ref{tab:activity}. 

\begin{table}[!h]
\centering
\begin{scriptsize}
\caption{Activities Used in Our Experiments}
\label{tab:activity}
\begin{tabular}{|l|l|l|}
\hline 
\multicolumn{1}{|c|}{\multirow{2}{*}{\textbf{Index}}} & \multicolumn{2}{c|}{\textbf{Recognizable Activities}} \\ \cline{2-3} 
\multicolumn{1}{|c|}{} & \multicolumn{1}{c|}{\textbf{Basic Activities}} & \multicolumn{1}{c|}{\textbf{Complex Activities}} \\ \hline \hline
\textbf{1} & Sitting & Taking Medicine \\ \hline
\textbf{2} & Standing & Eating \\ \hline
\textbf{3} & Lying& Watching TV \\ \hline
\textbf{4} & Walking & Reading magazine \\ \hline
\textbf{5} & Arm Movement & Cleaning table \\ \hline
\textbf{6} & Kicking & Vaccuming \\ \hline
%\textbf{7} & Bending & Bathing \\ \hline
\textbf{7} & Crouching & Toileting \\ \hline
\textbf{8} & Falling & Sleeping \\ \hline
%\textbf{10} & - & Making Coffee \\ \hline
%\textbf{11} & - & Taking Medicine \\ \hline
\end{tabular}
\end{scriptsize}
%\vspace{-4mm}
\end{table}

\begin{figure}[!tb]
\begin{center}
\begin{minipage}{4.5cm}
\includegraphics[width=4.5cm]{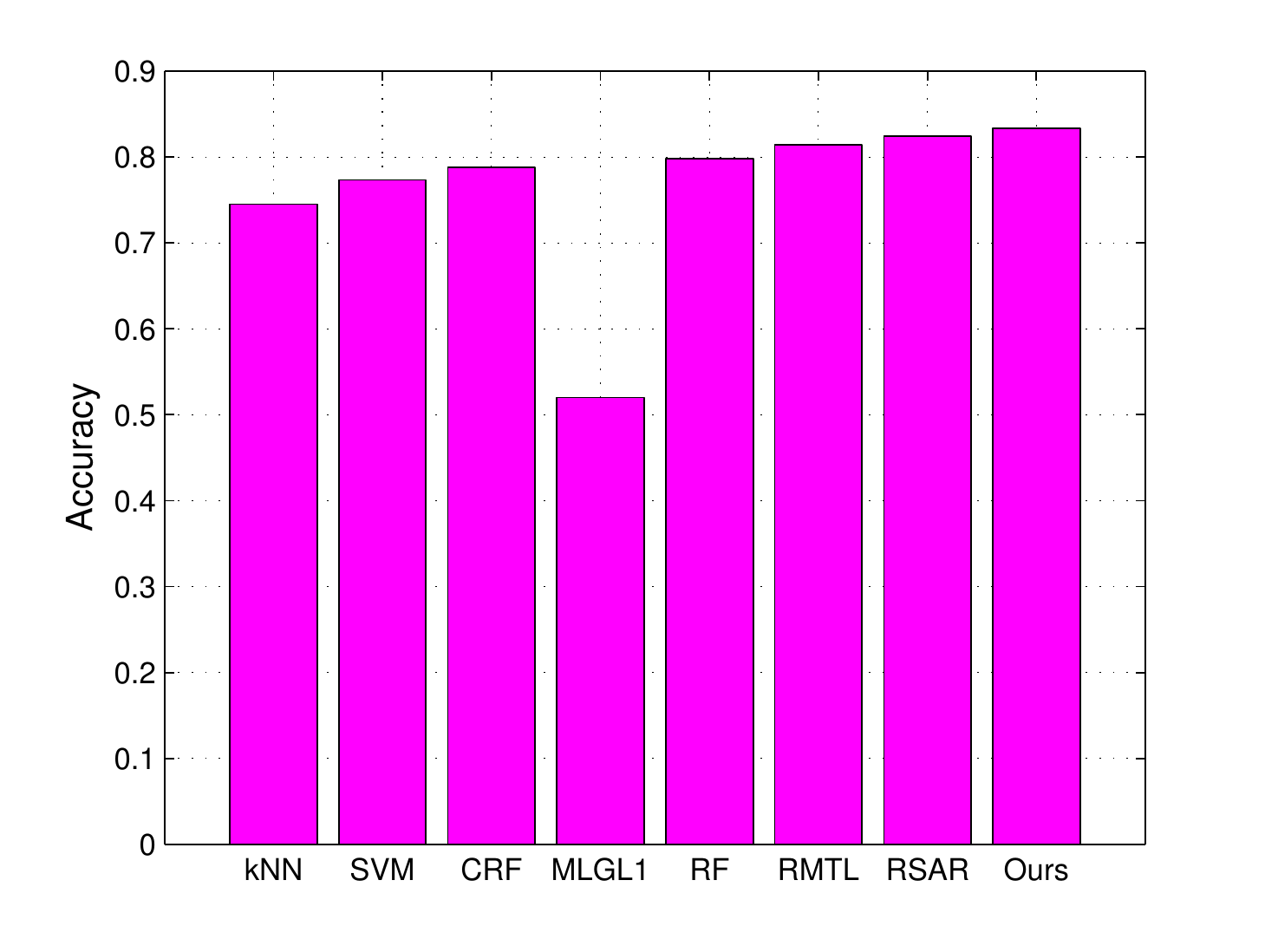}
\centering{(a)}
\end{minipage}
\begin{minipage}{4.5cm}
\includegraphics[width=4.5cm]{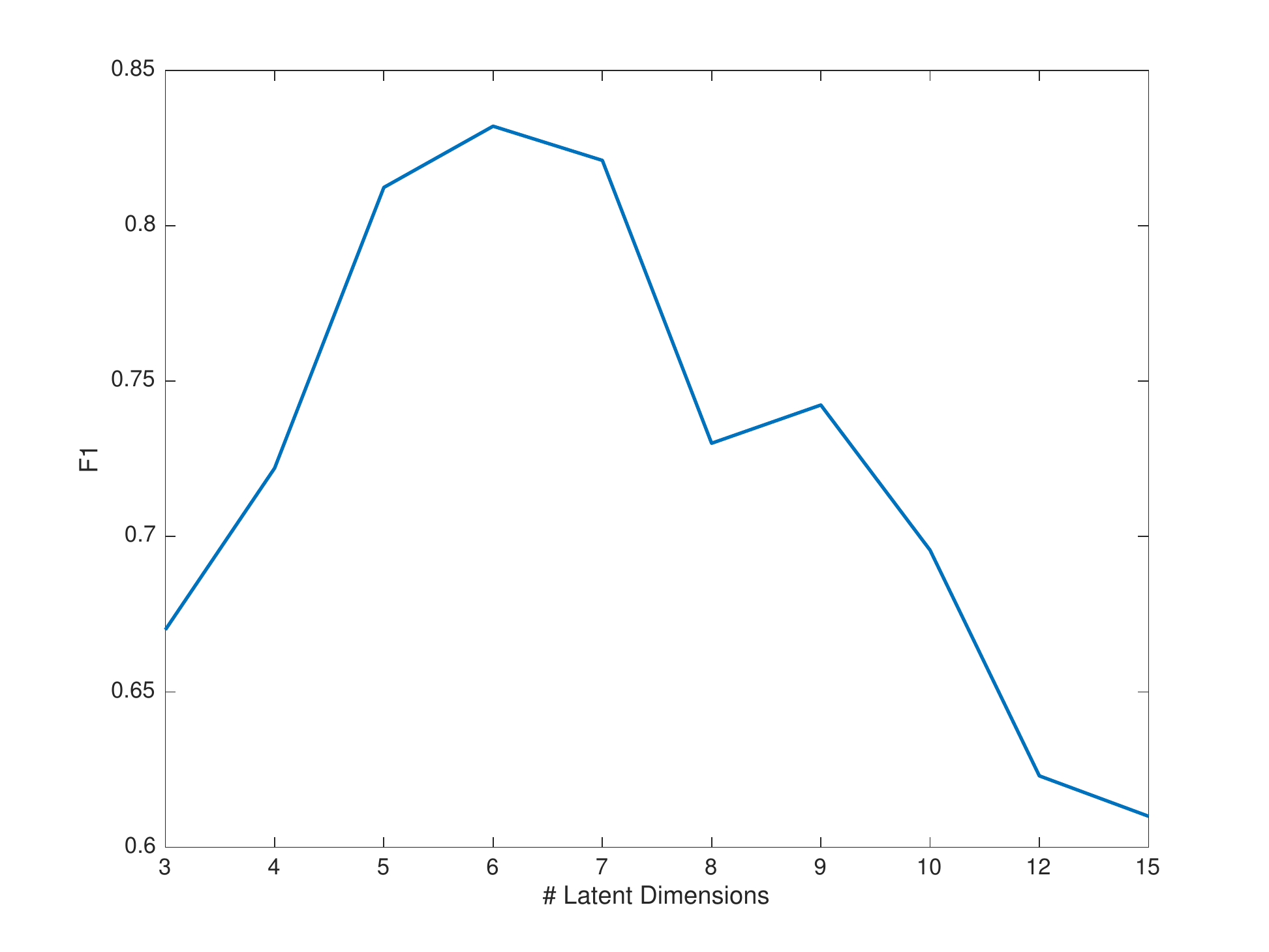}
\centering{(b)}
\end{minipage}
\begin{minipage}{4.5cm}
\includegraphics[width=4.5cm]{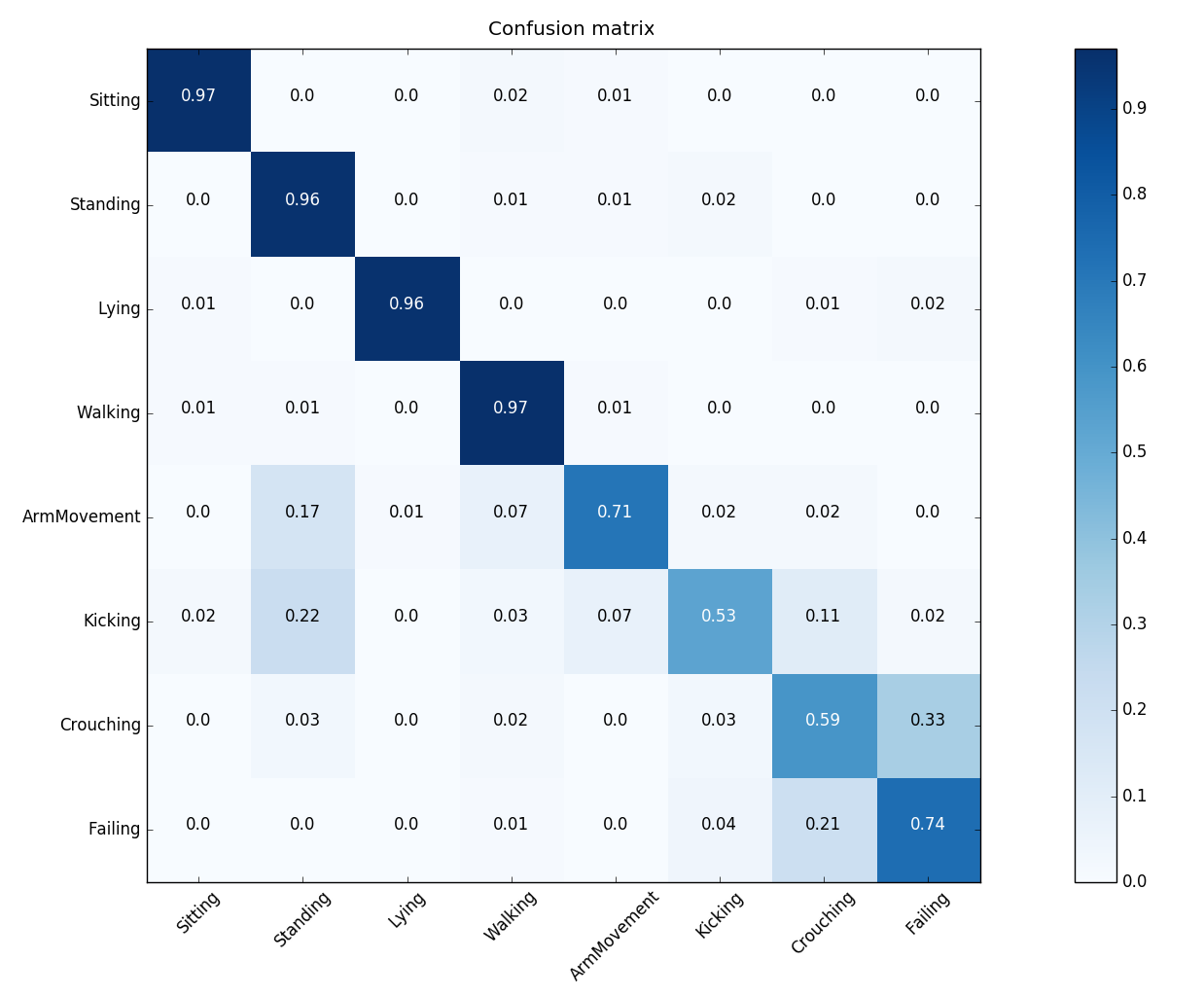}
\centering{(c)}
\end{minipage}
\begin{minipage}{4.5cm}
\includegraphics[width=4.5cm]{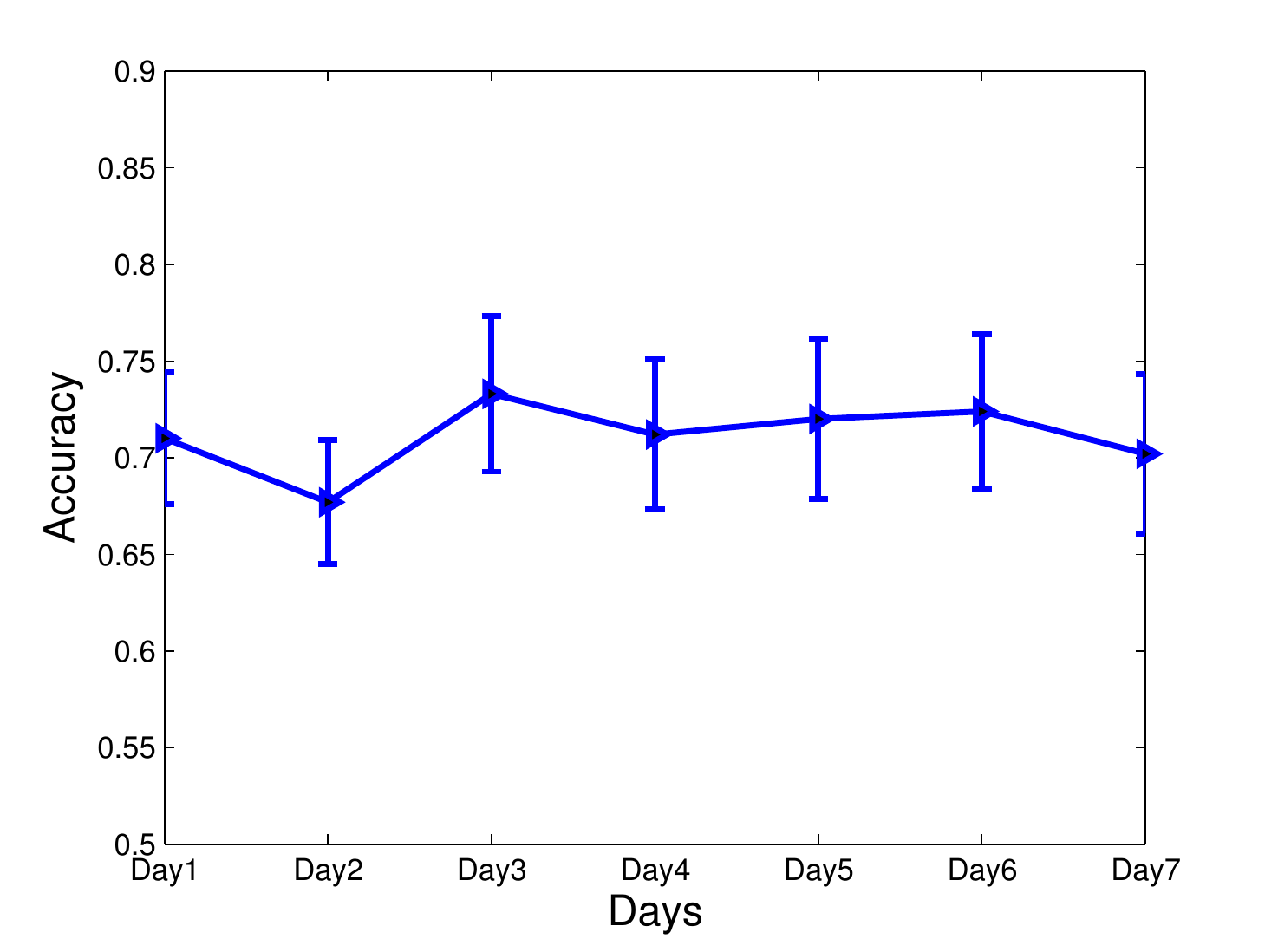}
\centering{(d)}
\end{minipage}
\caption{(a) Comparison with other methods (b) Performance comparison on different dimensions of shared subspace; (c) Confusion matrix; (d) High-level activity reasoning accuracy over 7 days.}
\label{fig:dictionarycompare}
\end{center}
\vspace{-5mm}
\end{figure}

We compare our method with a series of existing methods, including the widely-used sensor-based activity recognition methods such as  $k$ nearest neighbour ($k$NN), Support Vector Machine (SVM) with linear kernel, Conditional Random Field (CRF), Multinomial Logistic Regression with $\ell_1$ (MLGL1), Random Forest (RF), as well as some closely-related work such as RMTL \cite{chen2011integrating} (which integrates low-rank and group sparsity for multi-task learning) and RSAR~\cite{yao2016learning} (which integrates graph manifold learning and shared structure learning). We obtained the observations from the results as shown in Figure~\ref{fig:dictionarycompare}: 
\begin{itemize}
\item Our method outperforms all the compared methods (Figure \ref{fig:dictionarycompare}(a)). 
With larger dimensions of latent space $d$, the performance of our method degrades, especially when the size becomes bigger than $8$ (Figure \ref{fig:dictionarycompare}(b)). %We set $d = 5$ as the default value. 
\item The overall performance of the proposed approach is comparatively stable and consistent across all the activities (Figure~\ref{fig:dictionarycompare} (c)), despite the recognition accuracy presents slight ups and downs.
\item Combining with symbolic locations and object use events, the average recognition accuracy of high-level activities is generally stable at around 70\% (Figure~\ref{fig:dictionarycompare}(d)). 
\end{itemize}

We conclude this section with a brief discussion on 
%brief analysis of 
system latency, as fast activity detection and notification are critical for many applications, especially aged care applications. 
For example, we should send an alert to notify the caregivers as quickly as possible for medical assistance when a old person falls. Our system has $4 \sim 4.5$ seconds recognition latency, due to three main reasons: 
%i) 
\begin{itemize}
\item Our system evaluates subject's postures every 0.5 seconds using the latest $2$ seconds of the signal stream. In other words, if the current system time is at the timestamp $t$, our system will produce the predicted actions in the $[t-2,t-1]$ seconds, and $[t-1,t]$ seconds is used to backtrack check if the predicted label complies with predefined rules. For instance, assume that the label is estimated as: {\em lying in bed} at $[t-2,t-1]$ interval, if the predicted label in interval $[t-1,t]$ is {\em walking}, our system will determine the prediction as still {\em lying}. 
\item Signal collector is programmed with a timer to poll the signal variations with a predefined order of transmission, which takes around $1$ second to complete a new measurement with no workarounds. 
%iii) In addition, since 
\item Our system is integrated with a Web-based interface, which sends AJAX requests to services for the latest results %calculated by our algorithm modules 
and retrieve data fro databases to send back to the Web interface with updated DOM (document object model) elements. Completing such a querying process normally takes 300ms to 500ms. 
\end{itemize}

\section{Conclusion and Future Work}

In this paper, we propose the design and development of WITS, an automatic IoT-based in-home monitoring system. We propose a novel hybrid activity recognition framework that integrates multi-task based dictionary learning and rule-based reasoning 
%yield 
for better performance. An IoT middleware has been developed to provide an one-stop offer of
%better 
interoperability, scalability, and accessibility to the system. We have conducted extensive experiments to validate our proposed system and we anticipate our preliminary study and practical experience gained from this system can provide the foundation and inspiration for subsequent research on developing IoT applications for mainstream usage.

In the future, we will enhance our activity recognition approach by incorporating rich contexts for multi-person activity monitoring. As the evaluation in this paper is limited to a small testing scenario (i.e., the first author's home),
%A more promising and exciting direction is 
we plan to perform a larger scale study of the WITS system by working with an aged care center. 
%
%
%\vspace{2mm}
%For our future work, there are many challenges in improving the efficiency of the WITS system. One such challenge is concept drifting handling. The digital and physical worlds co-exist and interact simultaneously. Most things are resource-constrained, which are typically connected to the Web using lightweight, stateless protocols such as CoAP. We will further investigate novel methods for efficient transaction processing like collaborative sensing. 
% in our framework.  
%\section{Experiments}
%
%\subsection{Datasets}
%We validate and evaluate our proposed framework on two publicly available recognition datasets, and in a real living environment as well (discuss later). 
%
%
%
%\section{System Evaluation}
%\label{sec:experiment}

\bibliographystyle{abbrv}
%\bibliography{references,lina,linaicdm2015}
\bibliography{lina}
\end{document}